# The Glory of the Past and Geometrical Concurrency [*]

Cristian Prisacariu

Dept. of Informatics, Univ. of Oslo,
P.O. Box 1080 Blindern, N-0316 Oslo, Norway.
email: cristi@ifi.uio.no

November 1, 2018

## Abstract

This paper contributes to the general understanding of the *geometrical model of concurrency* that was named *higher dimensional automata* (HDAs) by Pratt and van Glabbeek. In particular we provide some understanding of the modal logics for such models and their expressive power in terms of the bisimulation that can be captured. The geometric model of concurrency is interesting from two main reasons: its generality and expressiveness, and the natural way in which autoconcurrency and action refinement are captured. Logics for this model, though, are not well investigated, where a simple, yet adequate, modal logic over HDAs was only recently introduced. As this modal logic, with two existential modalities, *during* and *after*, captures only split bisimulation, which is rather low in the spectrum of van Glabbeek and Vaandrager, the immediate question was what small extension of this logic could capture the more fine-grained *hereditary history preserving bisimulation* (hh)?

In response, the work in this paper provides several insights. One is the fact that the geometrical aspect of HDAs makes it possible to use for capturing the hh-bisimulation, a standard modal logic that does not employ event variables, opposed to the two logics (over less expressive models) that we compare with. The logic that we investigate here uses standard backward-looking modalities (i.e., past modalities) and extends the previously introduced logic (called HDML) that had only forward, action-labelled, modalities.

Since the direct proofs are rather intricate, we try to understand better the above issues by introducing a related model that we call *ST-configuration structures*, which extend the configuration structures of van Glabbeek and Plotkin. We relate this model to HDAs, and redefine and prove the earlier results in the light of this new model. These offer a different view on why the past modalities and geometrical concurrency capture the hereditary history preserving bisimulation. Additional correlating insights are also gained.

---

[*]Details for proofs of some of the results in this paper can be found in the on-line technical report [16].

**Tribute to Alan Turing.** Alan Turing was seeking beauty and naturalness in his work, e.g., with the Turing machine or the Turing test, and at the same time power of the theories and methods he was developing, e.g., with "the bombe". This is what makes a mathematical genius: beauty, power, and naturalness, that are constantly sought for when building methods for solving problems which most of us cannot even comprehend. Once these are achieved, then the problems and solutions open up like a book to the rest of us, and we read from "the book" (as Paul Erdős was saying) for many years to come. Our gratitude to Alan Turing can never be enough, compared to the benefits that we still get from his genius.

# 1 Introduction

Geometric concurrency was introduced by V. Pratt and R. van Glabbeek as the model of Higher Dimensional Automata two decades ago [9, 18]. This model of concurrency is more general and expressive than most other models like event structures or Petri nets (as studied by van Glabbeek [18]) and accommodates nicely action refinement, which is a good method for building systems from abstract specifications and refined to gradually more concrete implementations. Moreover, *HDAs* are not constrained to only before-after modelling and expose explicitly the choices in the system. It is a known issue in concurrency models that the combination of causality, concurrency, and choice is difficult; in this respect, *HDAs* and Chu spaces [3, 10] do a fairly good job [13].

Logics for this model, though, are not well investigated. A natural modal logic over HDAs was recently introduced as *HDML* in [14]. *HDML* contrasts with standard temporal/program logics in the fact that it can reason about *what holds "during" some concurrent events are executing*. A main question that this logic attracted was what bisimulation it captured. It turns out that this modal logic, with two existential modalities, *during* and *after*, captures only *split bisimulation*, cf. [15, Prop.4.3], which is rather low in the spectrum of van Glabbeek and Vaandrager [21]. This expressiveness lack is due to its forward-only modalities, that do not allow to look at the history of the concurrent execution. The natural question now is what small extension of this logic can capture the finest, hereditary history preserving bisimulation. This question becomes more interesting in the light of the fact that two recent logic developments [1, 7] study concurrency bisimulations (including hh) over concurrency models strictly less expressive than *HDAs*, yet using event-identifier variables inside more complex modalities.

For this, one concrete step in this paper is to define the *history-aware higher dimensional modal logic* (*hHDML*) which is a modal logic interpreted over higher dimensional automata (*HDAs*) that captures the *hereditary history-preserving bisimulation* (hh).

The *hHDML* logic that we present in this paper solves the question of what is a simple and natural extension of *HDML* that captures hh-bisimulation. Compared to related works, *hHDML* does not make reference to events, but talks only about labels; i.e., *hHDML* does not use event identifiers as in [7] or more complicated event-based modalities as in [1]. On the other hand, *hHDML* uses backward looking modalities as in [7] (whereas the feature of [1] is that it is a forward only logic). Moreover, *hHDML* uses single step modalities (both the forward and the backward), thus not looking at entire parts of the *HDA* model, as the modalities of [1, 7] do (or Until-like modalities of temporal logics).

The other feature of *hHDML* is that it is a modal logic defined over *HDA*, which are a model of true concurrency that is more expressive than the other models on which the logics of [1, 7] are defined (i.e., is more expressive than event structures or configuration structures). Therefore this makes it a good framework for investigating and comparing all these logics in a uniform manner; similar to what is done in [18] on comparing expressiveness of concurrency models by embedding them into *HDA* (i.e., defining the exact class of *HDA* that they capture); or similar to [5] where translations are made between logics that all capture the same bisimulation. Moreover, the simple syntax and natural interpretation of *hHDML* in the style of standard modal logic makes it a good candidate for investigating modal characterizations of the spectrum of true concurrency bisimulations [8, 19, 21], as was done with modal logics for standard process equivalences in [17] and recently using event identifier modalities in [1].

In the second part of this paper we want to convey a better understanding of why *hHDML* over *HDA* can capture hh-bisimulation without referring to event variables. On top, we want to get more tight correlations between this logic and *HDAs*, and the related logics over the event structures and configuration structures, as well as between the definitions of bisimulations over *HDA* and over the other partial order models, cf. [19].

For this we introduce *ST-configuration structures*, which are a natural extension of configuration structures to the setting of higher dimensional automata. Configuration structures [20] are used in [19] as the most expressive model of concurrency which has a natural way of defining refinement and the partial order bisimulations. The notion of an ST-configuration has been used in [21] to define ST-bisimulation and in [18] in the context of *HDA*. But the model of *ST-configuration structures*, as we define here for capturing concurrency, does not appear elsewhere. We think that a main characteristic of higher dimensional automata is captured by ST-configuration structures, opposed to the standard configuration structures; this is the power to look at the currently executing concurrent events. At the level of the modal logic *HDML* this is the power to talk about what happens *during* the concurrent execution of one or more events. This is opposed to standard modal logics that talk only about what happens *after* the execution of one or more events.

For ST-configuration structures we show how they are a natural extension of configuration structures and define related notions that steam from the later. We also define notions of steps and paths and define the bisimulations in this new context. We show how these relate to existing bisimulations for the other models. For this we first relate ST-configuration structures also to *HDAs* by identifying the corresponding class of ST-configuration structures, with the particular property of adjacent-closure. We also define the class of stable ST-configuration structures and relate this with their counterpart in stable configuration structures. The classical notions of concurrency, causality, and conflict are not interrelated as in the case of event structures or configuration structures; but are more loose, as is with *HDAs*.

## 2  Preliminaries on Higher Dimensional Automata

In this section we define higher dimensional automata (*HDA*) following the terminology of [13, 18]. We also define additional notions like paths, adjacency, bisimulations, and the restriction to acyclic and cubical *HDAs*.

For an intuitive understanding of the *HDA* model consider the standard example [13, 18] pictured in Figure 1. It represents a *HDA* that models two concurrent events which are labelled by $a$ and $b$ (we can also have the same label $a$ for both events). The *HDA* has four states, $q_0^1$ to $q_0^4$, and four transitions between them. This would be the standard picture for interleaving, but in the case of *HDA* there is also a square $q_2$. Traversing through the interior of the square means that both events are executing. When traversing on the lower transition means that event one is executing but event two has not started yet, whereas, when traversing through the upper transition it means that event one is executing and event two has finished already. In the states there is no event executing; in particular, in state $q_0^3$ both events have finished, whereas in state $q_0^1$ no event has started yet.

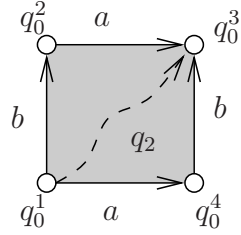

Figure 1: Example of a *HDA* with two concurrent events.

Similarly, *HDAs* allow to represent three concurrent events through a cube, or more events through hypercubes. Causality of events is modelled by sticking such hypercubes one after the other. For our example, if we omit the interior of the square (i.e., the grey $q_2$ is removed) we are left with a description of a system where there is the choice between two sequences of two events, i.e., $a;b + b;a$.

**Definition 2.1 (higher dimensional automata)** *A cubical set $H = (Q, \overline{s}, \overline{t})$ is formed of a family of sets $Q = \bigcup_{n=0}^{\infty} Q_n$ with all sets $Q_n$ disjoint, and for each $n$, a family of maps $s_i, t_i : Q_n \to Q_{n-1}$ with $1 \leq i \leq n$ which respect the following cubical laws:*

$$\alpha_i \circ \beta_j = \beta_{j-1} \circ \alpha_i, \quad 1 \leq i < j \leq n \text{ and } \alpha, \beta \in \{s, t\}. \tag{1}$$

*In $H$, the $\overline{s}$ and $\overline{t}$ denote the collection of all the maps from all the families (i.e., for all $n$). A higher dimensional structure $(Q, \overline{s}, \overline{t}, l)$ over an alphabet $\Sigma$ is a cubical set together with a labelling function $l : Q_1 \to \Sigma$ which respects $l(s_i(q)) = l(t_i(q))$ for all $q \in Q_2$ and $i \in \{1, 2\}$. A higher dimensional automaton $(Q, \overline{s}, \overline{t}, l, I, F)$ is a higher dimensional structure with two designated sets of* initial *and* final *cells $I \subseteq Q_0$ and $F \subseteq Q_0$.*

We call the elements of $Q_0, Q_1, Q_2, Q_3$ respectively *states*, *transitions*, *squares*, and *cubes*, whereas the general elements of $Q_n$ are called *cells* (also known as n-cell, n-dimensional cubes, or hypercubes). For a transition $q \in Q_1$ the $s_1(q)$ and $t_1(q)$ represent respectively its source and its target cells (which are *states* from $Q_0$ in this case). Similarly for a general n-cell $q \in Q_n$ there are $n$ source cells and $n$ target cells all of dimension $n-1$. Intuitively, an n-dimensional cell $q$ represents a configuration of a concurrent system in which $n$ events are performed at the same time, i.e., concurrently. A source cell $s_i(q)$ represents the configuration of the system before the starting of the $i^{th}$ event, whereas the target cell $t_i(q)$ represents the configuration of the system immediately after the termination of the $i^{th}$ event. A transition of $Q_1$ represents a configuration of the system in which a single event is being performed.

The cubical laws account for the *geometry* (concurrency) of the *HDAs*; there are four kinds of cubical laws depending on the instantiation of $\alpha$ and $\beta$. For the example of Figure 1 consider the cubical law where $\alpha$ is instantiated to $t$ and $\beta$ to $s$, and $i = 1$ and $j = 2$: $t_1(s_2(q_2)) = s_1(t_1(q_2))$. In the left hand side, the second source cell of $q_2$ is, in this case, the transition $s_2(q_2) = q_1^1 = (q_0^1, q_0^2)$ and the first target cell of $q_1^1$ is $q_0^2$ (the only target cell because $s_2(q_2) \in Q_1$); this must be the same cell when taking the right hand side of the cubical law, i.e., the first target cell is $t_1(q_2) = q_1^2 = (q_0^2, q_0^3)$ and the first source of $q_1^2$ is $q_0^2$.

The terminology that we adopt now steams from the Chu spaces representation of *HDAs* and triadic event structures [11, 13]. A cell $q_n$ of dimension $n$ represents $n$ events happening at the same time. Therefore we assume a set $E$, with $|E| = n$, which for our purposes denotes the concurrent events. In the *HDA* case each event may be in three phases: *not started*, *executing*, and *terminated* (as opposed to only terminated or not started in the standard event oriented models). In consequence, we associate one valuation $E \to \{0, \frac{1}{2}, 1\}$, where $\frac{1}{2}$ means executing, to each cell in the hypercube $q_n = 3^E$. A hypercube (or n-cell) $q_n$ is formed of $3^n$ cells, which we call *faces of $q_n$*; these are those cells on lower levels that are connected to $q_n$ through $s$ or $t$ maps ($F(q_n) = \cup_{0 \leq k \leq n} F^k(q_n)$ where $F^n = \{q_n\}$ and $F^k(q_n) = \{q \in Q_k \mid (q = s_i(q_{k+1})$ or $q = t_i(q_{k+1}))$ and $q_{k+1} \in F^{k+1}\}$). In the context of a single hypercube $3^E$ we denote the cells of the cube by lists of $|E|$ elements $e_1 e_2 \ldots e_{|E|}$ where each $e_i$ takes values in $\{0, \frac{1}{2}, 1\}$ and represents the status of the $i^{th}$ event of the *HDA*. The dimension of a cell in this hypercube is equal to the number of $\frac{1}{2}$ in its corresponding valuation. With the above conventions, the cells of dimension 0 (i.e., the states of the hypercube $q_n$) are denoted by the corresponding valuation restricted to only the two values $\{0, 1\}$.

**Definition 2.2 (general labelling)** *Because of the condition $l(s_i(q)) = l(t_i(q))$ for all $q \in Q_2$, all the edges $e_1 \ldots e_{i-1} \frac{1}{2} e_{i+1} \ldots e_{|E|}$, with $e_j \in \{0, 1\}$ for $j \neq i$, have the same label. Denote this as the label $l_i$. The label of a general cell $q \in Q_n$ is the multiset of $n$ labels $l_{j_1} \ldots l_{j_n}$ where the $j$'s are exactly those indexes in the representation of $q$ for which $e_j$ has value $\frac{1}{2}$.*

**Definition 2.3 (paths in *HDAs*)** *A single step in a HDA is either $q_{n-1} \xrightarrow{s_i} q_n$ with $s_i(q_n) = q_{n-1}$ or $q_n \xrightarrow{t_i} q_{n-1}$ with $t_i(q_n) = q_{n-1}$, where $q_n \in Q_n$ and $q_{n-1} \in Q_{n-1}$ and $1 \leq i \leq n$. A path $\pi \triangleq q^0 \xrightarrow{\alpha^1} q^1 \xrightarrow{\alpha^2} q^2 \xrightarrow{\alpha^3} \ldots$ is a sequence of single steps $q^j \xrightarrow{\alpha^{j+1}} q^{j+1}$, with $\alpha^j \in \{s_i, t_i\}$. We say that $q \in \pi$ iff $q = q^j$ appears in one of the steps in $\pi$. The first cell in a path is denoted $st(\pi)$ and the ending cell in a finite path is $en(\pi)$. We are interested in the observable content of a path which is the sequence of annotated labels of the single steps, hence we denote a single step $q_{n-1} \xrightarrow{l_i(q_n)^+} q_n$ when $q_{n-1} = s_i(q_n)$ and $q_n \xrightarrow{l_i(q_n)^-} q_{n-1}$ when $q_{n-1} = t_i(q_n)$. We denote by $\pi \xrightarrow{a^\pm} \pi'$ when the path $\pi'$ extends $\pi$ by a single step labelled by $a \in \Sigma$; the step may be either a start or a terminate step.*

Many of the results in this paper work with *acyclic* and *cubical HDAs* in the following sense. Such *HDAs* are the ones usually considered in the literature on concurrent systems and are more general than most of the true concurrency models [13, 18].

**Definition 2.4 (acyclic and cubical *HDAs*)** *A HDA is called* acyclic *if no path visits a cell twice (i.e., no path results in cycles). An acyclic HDA is called* cubical *if for any cell all its faces are different.*

Acyclic and cubical *HDAs* result from the definition of *HDA* involving cubical complexes, and hence, the concurrent system is built by putting together cubes of varying dimensions to share some of their faces. The example of the empty square is built from cubes of dimension 1 that share their end point faces. The same can be said about all the examples that we give in Section 3.1. A different view, that may be more appealing to a more abstract mind with aesthetic inclinations, is as "sculpting out" (as Pratt called it) from a big hypercube formed of all the events in the system (not just those that are concurrent, as is the case with a single n-cell), those of its faces that are not relevant. In the same example from the beginning, we would take a 2-cell and carve out its interior face, i.e., from level $Q_2$.

**Definition 2.5 (adjacency and homotopy)** *Two finite paths $\pi$ and $\pi'$ are called l-adjacent, denoted $\pi \xleftrightarrow{l} \pi'$, if one can be obtained from the other by replacing the segment $\xrightarrow{\alpha^l} q^l \xrightarrow{\alpha^{l+1}}$ in one of the following four ways (both directions are allowed), where the path $\pi = q^0 \xrightarrow{\alpha^0} q^1 \xrightarrow{\alpha^1} q^2 \xrightarrow{\alpha^2} \ldots q^m$, with $l < m$:*

- *replace segments $\xrightarrow{s_i} q \xrightarrow{s_j}$ and $\xrightarrow{s_{j-1}} q' \xrightarrow{s_i}$, or $\xrightarrow{t_j} q \xrightarrow{t_i}$ and $\xrightarrow{t_i} q' \xrightarrow{t_{j-1}}$, or*

- *$\xrightarrow{s_i} q \xrightarrow{t_j}$ and $\xrightarrow{t_{j-1}} q' \xrightarrow{s_i}$, or $\xrightarrow{s_j} q \xrightarrow{t_i}$ and $\xrightarrow{t_i} q' \xrightarrow{s_{j-1}}$.*

*where $i < j$.* Homotopy *is the reflexive and transitive closure of adjacency. Two* homotopic *paths share both their end points. The* homotopy class *of a cell $q$ is the set of all homotopic paths that end in $q$ (and thus start in the initial cell). This is the* history *of $q$.*

**Corollary 2.6 (cf. [18, sec.7.5])** *For a path $\pi$ and a point $l > 1$ there exists a unique path $\pi'$ that is l-adjacent with $\pi$.*

**Definition 2.7 (hh-bisimulation)** *Two higher dimensional automata $(\mathcal{H}_A, q_A^0)$ and $(\mathcal{H}_B, q_B^0)$ (with $q_A^0$ and $q_B^0$ two initial cells) are* hereditary history-preserving bisimulation equivalent *(hh-bisimilar) if there exists a binary relation $R$ between their paths starting at $q_A^0$ respectively $q_B^0$ that respects the following:*

1. *if $\pi_A R \pi_B$ and $\pi_A \xrightarrow{a^{\pm}} \pi'_A$ then $\exists \pi'_B$ with $\pi_B \xrightarrow{a^{\pm}} \pi'_B$ and $\pi'_A R \pi'_B$;*

2. *if $\pi_A R \pi_B$ and $\pi_B \xrightarrow{a^{\pm}} \pi'_B$ then $\exists \pi'_A$ with $\pi_A \xrightarrow{a^{\pm}} \pi'_A$ and $\pi'_A R \pi'_B$;*

3. *if $\pi_A R \pi_B$ and $\pi_A \xleftrightarrow{l} \pi'_A$ then $\exists \pi'_B$ with $\pi_B \xleftrightarrow{l} \pi'_B$ and $\pi'_A R \pi'_B$;*

4. *if $\pi_A R \pi_B$ and $\pi_B \xleftrightarrow{l} \pi'_B$ then $\exists \pi'_A$ with $\pi_A \xleftrightarrow{l} \pi'_A$ and $\pi'_A R \pi'_B$;*

5. *if $\pi_A R \pi_B$ and $\pi'_A \xrightarrow{a^{\pm}} \pi_A$ then $\exists \pi'_B$ with $\pi'_B \xrightarrow{a^{\pm}} \pi_B$ and $\pi'_A R \pi'_B$;*

6. *if $\pi_A R \pi_B$ and $\pi'_B \xrightarrow{a^{\pm}} \pi_B$ then $\exists \pi'_A$ with $\pi'_A \xrightarrow{a^{\pm}} \pi_A$ and $\pi'_A R \pi'_B$.*

*Denote this as $(\mathcal{H}_A, q_A^0) \overset{hh}{\sim} (\mathcal{H}_B, q_B^0)$.*

# 3  History-aware Higher Dimensional Modal Logic

We extend the higher dimensional modal logic of [14] with backward looking modalities in the style of past temporal logics [6] or PDL with converse [4, ch.10.5]. Call this extension *history-aware higher dimensional modal logic*, for short *hHDML*. This logic follows the tradition and style of standard modal languages [2]. *hHDML* is a multi-modal logic with the modalities labelled from a finite set of action labels, which are the same $\Sigma$ labels of the *HDA* that we interpret the logic over.

**Definition 3.1 (history-aware higher dimensional modal logic)** *A formula $\varphi$ in the language of hHDML is constructed using the grammar below, from a set $\Phi_B$ of atomic propositions, with $\phi \in \Phi_B$, which are combined using the propositional symbols $\bot$ and $\rightarrow$ (from which all other standard propositional operations are generated), and using the forward modalities $\{a\}$ and $\langle a \rangle$ and backward modalities $\overline{\{a\}}$ and $\overline{\langle a \rangle}$, all parametrized by the action labels in $\Sigma$.*

$$\varphi \quad := \quad \phi \mid \bot \mid \varphi \rightarrow \varphi \mid \{a\}\varphi \mid \langle a \rangle \varphi \mid \overline{\{a\}}\varphi \mid \overline{\langle a \rangle}\varphi$$

We call $\{a\}$ the *during modality* and $\langle a \rangle$ the *after modality*. The intuitive reading of $\{a\}\varphi$ is: "pick some event from the ones currently not running (must exist at least one not running) and start it; in the new configuration of the system (during which, one more event is executing) the formula $\varphi$ must hold". The intuitive reading of $\langle a \rangle \varphi$ is: "pick some event from the ones currently running concurrently (must exist one running) and terminate it; in the new configuration of the system the formula $\varphi$ must hold".

These two modalities only make the higher dimensional modal logic of [14]. *hHDML* adds the two new backward looking modalities $\overline{\{a\}}$ and $\overline{\langle a \rangle}$, increasing the expressive power of *HDML* to the point that it captures the hh-bisimulation (as we prove in this section). Intuitively, if the forward modalities $\{a\}$ and $\langle a \rangle$ were following a path, the past modalities $\overline{\{a\}}$ and $\overline{\langle a \rangle}$ are walking on paths backwards (from right to left), undoing events that may have been started or terminated.

The models of *hHDML* are higher dimensional structures together with a valuation function $\mathcal{V} : Q \rightarrow 2^{\Phi_B}$ which associates a set of atomic propositions to each cell (of any dimension). This means that $\mathcal{V}$ assigns some propositions to each state of dimension 0, to each transition of dimension 1, to each square of dimension 2, to each cube of dimension 3, etc. Denote a model of *hHDML* by $\mathcal{H} = (Q, \overline{s}, \overline{t}, l, \mathcal{V})$. A *hHDML* formula is evaluated in a cell of such a model $\mathcal{H}$.

**Definition 3.2 (satisfiability)** *Table 1 defines recursively the satisfaction relation $\models$ of a formula $\varphi$ w.r.t. a model $\mathcal{H}$ in a particular n-cell q (for some arbitrary n); denote this as $\mathcal{H}, q \models \varphi$. The notions of satisfiability and validity are defined as usual.*

The four modalities have an existential flavour. The universal correspondents are defined in the standard style of modal logic. We denote these modalities using square brackets; i.e., respectively $[\![a]\!]\varphi$, $[a]\varphi$, $\overline{[\![a]\!]}\varphi$ and $\overline{[a]}\varphi$. The intuitive reading of $\overline{[\![a]\!]}\varphi$ is: "$\varphi$ holds in all those configurations of the system from which the current configuration can be reached by starting some event labelled by $a$". In other words, it looks at all possible ways of undoing the start of some event labelled with $a$.

**Definition 3.3 (modal equivalence)** *Define the hHDML modal equivalence as the relation $\stackrel{hHDML}{\sim}$ between cells, s.t.:*

$$(\mathcal{H}, q) \stackrel{hHDML}{\sim} (\mathcal{H}', q') \text{ iff } \forall \varphi : \mathcal{H}, q \models \varphi \Leftrightarrow \mathcal{H}', q' \models \varphi.$$

$\mathcal{H}, q \models \phi$      iff $\phi \in \mathcal{V}(q)$.

$\mathcal{H}, q \not\models \bot$

$\mathcal{H}, q \models \varphi_1 \to \varphi_2$ iff when $\mathcal{H}, q \models \varphi_1$ then $\mathcal{H}, q \models \varphi_2$.

$\mathcal{H}, q \models \{a\}\varphi$    iff assuming $q \in Q_n$ for some $n$, $\exists q' \in Q_{n+1}$ s.t.
$s_i(q') = q$ for some $1 \leq i \leq n+1$, $l(q') = l(q)a$ and $\mathcal{H}, q' \models \varphi$.

$\mathcal{H}, q \models \langle a \rangle \varphi$    iff assuming $q \in Q_n$ for some $n$, $\exists q' \in Q_{n-1}$ s.t.
$t_i(q) = q'$ for some $1 \leq i \leq n$, $l(q) = l(q')a$ and $\mathcal{H}, q' \models \varphi$.

$\mathcal{H}, q \models \overline{\{a\}}\varphi$    iff assuming $q \in Q_n$ for some $n$, $\exists q' \in Q_{n-1}$ s.t.
$s_i(q) = q'$ for some $1 \leq i \leq n$, $l(q) = l(q')a$ and $\mathcal{H}, q' \models \varphi$.

$\mathcal{H}, q \models \overline{\langle a \rangle}\varphi$    iff assuming $q \in Q_n$ for some $n$, $\exists q' \in Q_{n+1}$ s.t.
$t_i(q') = q$ for some $1 \leq i \leq n+1$, $l(q') = l(q)a$ and $\mathcal{H}, q' \models \varphi$.

Table 1: Semantics for *hHDML*.

Two models $\mathcal{H}_A$ and $\mathcal{H}_B$ are modally equivalent iff their initial cells are; i.e., iff $(\mathcal{H}_A, I_A) \overset{hHDML}{\sim} (\mathcal{H}_B, I_B)$.

For proving the next results we need to relate the equivalence of cells, as in the modal equivalence, with the equivalence of paths, as in the hh-bisimulation.

We say the two paths of two *HDAs* are modal equivalent iff they have the same length and all their corresponding cells are modal equivalent.

**Proposition 3.4 (*hHDML* captures hh-bisimulation)** *The relations $\overset{hHDML}{\sim}$ and $\overset{hh}{\sim}$ coincide.*

**Proof:** The proof that $\overset{hh}{\sim}$ is in $\overset{hHDML}{\sim}$ is simpler, using *reductio ad absurdum* to assume a formula that holds on one model but not on the other, and then employ induction on the structure of the formula. Use the forward steps of $\overset{hh}{\sim}$ when we work with the forward modalities, and the back steps (5 and 6) for backward modalities.

The proof that $\overset{hHDML}{\sim}$ is in $\overset{hh}{\sim}$ is the more involved part. It shows that $\overset{hHDML}{\sim}$ respects the six rules of being a $\overset{hh}{\sim}$ over paths. For the forward and backward rules (1,2 and 5,6) we employ the assumptions of image-finite (i.e., finite choices) and finite concurrency, using *reductio ad absurdum* to construct a formula that will contradict the initial assumptions. The adjacency rules (3 and 4) are more involved. Knowing that for each of the two initial paths there exists a unique *l*-adjacent path (cf. Corollary 2.6) the proof reduces to showing that these two are modal equivalent, i.e., that their respective cells satisfy the same formulas. We do this for each of the four *l*-adjacency replacements, using induction on *l* and on the dimension of the cell involved in the *l*-adjacency replacement. We also use the proof principle *reductio ad absurdum* and show that whenever assuming an "error" in the $\overset{hh}{\sim}$ that we are building using $\overset{hHDML}{\sim}$, we can change the relation so far by interlacing cells (or indexes of maps), so that the *l*-adjacency is respected. □

## 3.1 Expressiveness of *hHDML* through examples

We give an intuition for the expressive power of *hHDML* logic interpreted over *HDAs* by exemplifying its distinguishing power wrt. the modal equivalence.

For each of the two systems that are compared in each example we give two presentations: one in the style of event structures, and the other as a *HDA*. The event structure style (using partial

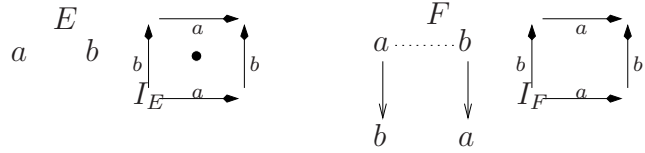

Figure 2: Concurrency vs. interleaving. $E \models \{a\}\{b\}\top$, $F \not\models \{a\}\{b\}\top$.

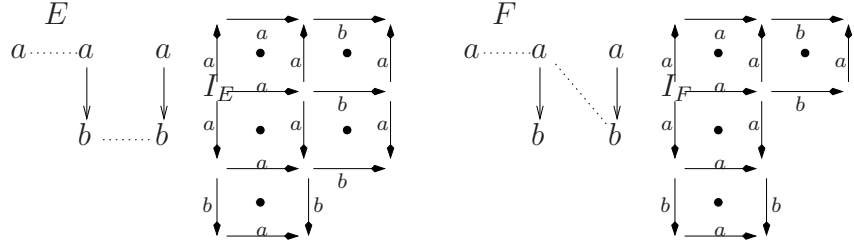

Figure 3: The limits of weak history preserving bisimulation. $I_E \models \{a\}\{a\}[\![a]\!]\{b\}\top$, $I_F \not\models \{a\}\{a\}[\![a]\!]\{b\}\top$

orders) is displayed on the left using just labels (a,b,...) for the events and arrows (downwards) denoting causality between the events, and dotted lines denoting conflict. We omit arrows and lines that can be inferred through transitive closure, and two events that are not connected by any line or arrow are meant as concurrent. The *HDA* presentation is displayed on the right using squares. We represent by a black dot the fact that a square is filled in. A side of a square (i.e., a transition, or cell of dimension 1) is represented through a different arrow showing the direction of the space, to account for the fact that we omit the *s* and *t* maps. We label these arrows with the corresponding action label of the corresponding event that the 1-cell is supposed to model. All the examples, except the first two, cannot be distinguished only using the initial *HDML* logic; they require the past modalities.

**Example 3.5 (concurrency vs. interleaving)** *This is the standard example of how HDAs distinguish between concurrency and interleaving, and how the original higher dimensional modal logic would distinguish these two. In Figure 2, the system $E$ represents $a\|b$ whereas system $F$ represents $a;b+b;a$. The two HDA presentations are distinguished by the hHDML formula*

$$\varphi \triangleq \{a\}\{b\}\top$$

*which holds on $E$ but not on $F$ in their respective initial cells.*

**Example 3.6 (limits of wh-bisimulation)** *This example is from [19, ex.9.3] and shows the limits of wh-bisimulation, which cannot distinguish the two systems; but they can be distinguished by pomset-bisimulation. The two systems of this example are depicted in their partial order and HDA presentations in Figure 3. For each system, the arrow in the HDA presentation that is labelled with a and goes out of the initial cell horizontally, corresponds to the event in the upper right corner of the partial order presentation. These two HDA systems are distinguished by the hHDML formula*

$$\varphi \triangleq \{a\}\{a\}[\![a]\!]\{b\}\top$$

*which holds on $E$ but not on $F$ in their respective initial cells.*

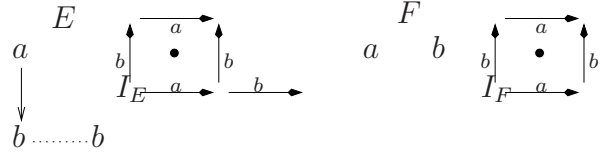

Figure 4: ST vs. causality based bisimulations. $I_E \models \{a\}\langle a\rangle\{b\}\neg\overline{\langle a\rangle}\top$, $I_F \not\models \{a\}\langle a\rangle\{b\}\neg\overline{\langle a\rangle}\top$

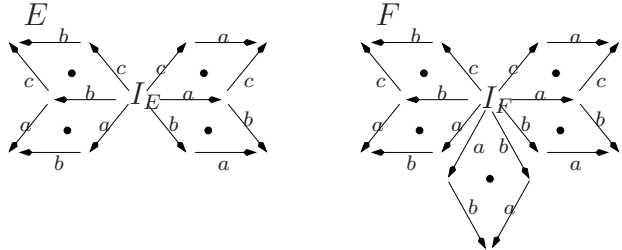

Figure 5: Absorbtion law example systems: $E = (a|(b+c)) + (b|(a+c))$ and $F = (a|(b+c)) + a|b + (b|(a+c))$.

**Example 3.7 (causality vs. ST-bisimulations)** *This example from [21, sec.5.4] shows the difference between the ST-bisimulation and the causality based bisimulations like pomset or h-bisimulation. The two systems of this example are depicted in their partial order and HDA presentations in Figure 4. These two systems are indistinguishable by the ST-bisimulation but are distinguished by any of the causality based bisimulations because one needs to look at the history, i.e., the homotopy class of a cell. The two HDA presentations are distinguished by the hHDML formula*

$$\varphi \triangleq \{a\}\langle a\rangle\{b\}\neg\overline{\langle a\rangle}\top$$

*which holds on $E$ but not on $F$ in their respective initial cells.*

**Example 3.8 (absorbtion law)** *The* absorbtion law example *is used in [1, 19] to show that hh-bisimulation has strictly more distinguishing power than h-bisimulation; where in [7] it is shown that the two examples are distinguished already by hwh-bisimulation. The two systems of this example are:*

$$E = (a|(b+c)) + (b|(a+c)) \text{ and } F = (a|(b+c)) + a|b + (b|(a+c)),$$

*as depicted in their HDA presentation in Figure 5. The horizontal right arrow labelled by $a$ going out of either $I_E$ or $I_F$ corresponds to the left-most $a$-labelled event in their respective CCS description. There are no dependencies, only conflicting relations (expressed using the CCS $+$). These two HDA systems are distinguished by the hHDML formula*

$$\varphi \triangleq [\![a]\!][\![b]\!](\overline{\{b\}}\{c\}\top \vee \overline{\{a\}}\{c\}\top)$$

*which holds on $E$ but not on $F$ in their respective initial cells.*

**Example 3.9 (conflicting futures)** *This example from [1, ex.3] is meant there to show the need for quantification over event variables in conflict with previously bound events. The two systems of this example are depicted in their partial order and HDA presentations in Figure 6. These two systems are indistinguishable by the h-bisimulation but are distinguished by hh-bisimulation. The*

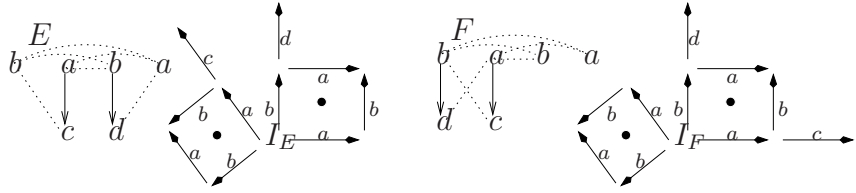

Figure 6: Conflicting futures of [1]. $\{I_E \models, I_F \not\models\}\ [\![a]\!][\![b]\!](\langle b \rangle \overline{\{a\}}\{d\}\top \vee \langle a \rangle \overline{\{b\}}\{c\}\top)$

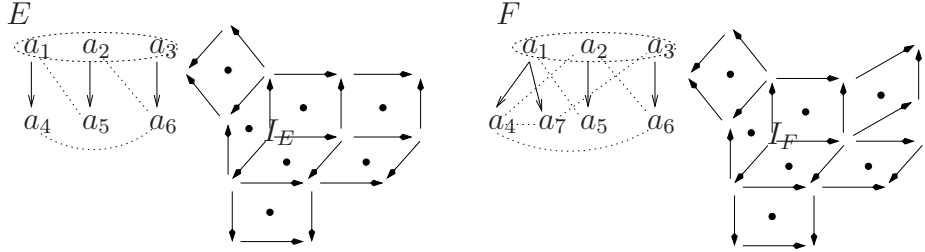

Figure 7: Example systems for non-binary conflict and event identifiers.

*horizontal right arrow going out of $I_E$ labelled with $a$ corresponds to the right-most $a$-labelled event in $E$. The two HDA presentations are distinguished by the hHDML formula*

$$\varphi \triangleq [\![a]\!][\![b]\!](\langle b \rangle \overline{\{a\}}\{d\}\top \vee \langle a \rangle \overline{\{b\}}\{c\}\top)$$

*which holds on $E$ but not on $F$ in their respective initial cells.*

**Example 3.10 (non-binary conflict and the need for event identifiers)** *This example is taken from [7, ex.4.8] where it is meant to illustrate that the event identifier logic indeed needs the event variables in order to make the distinction between the two concurrent systems which are distinguishable by the hh-bisimulation (the other examples of that paper can be distinguished by the logic without the need of event variables). The hHDML logic can distinguish these systems (thus, without the need of event variables). Another line that this example draws is that the HDA model can naturally express non-binary conflicting events. This means going beyond the expressiveness of prime event structures. Note that the examples of [19] are all using binary conflict. These two HDA systems of Figure 7 are distinguished by the hHDML formula*

$$\varphi \triangleq \{a\}\{a\}([a]\{a\}\top \wedge \langle a \rangle\{a\}\overline{\{a\}}\{a\}\langle a \rangle \overline{\{a\}}\langle a \rangle \neg \{a\}\top)$$

*which holds on $E$ but not on $F$ in their respective initial cells.*

## 4 ST-configuration structures

For a better understanding of the previous results we develop here the theory of *ST-configuration structures*. We do this along the lines drawn by Pratt [12,13], extending the configuration structures of van Glabbeek and Plotkin [20] starting from the notion of ST-configuration as defined in [18,21]. ST-configurations structures are a natural generalization of configuration structures to the setting of *HDA*. In this section we translate and relate to notions from *HDA* in ST-configuration structures, thus showing strong correlations between the two and eventually explaining better why the *hHDML* over *HDAs* captures hh-bisimulation without requiring quantification over event variables, i.e., only with the standard modal logic machinery.

**Definition 4.1 (ST-configuration)** *An ST-configuration is a pair of sets $(S,T)$ with the following property:*

$$\text{(start before terminate)} \quad T \subseteq S.$$

Intuitively $S$ contains the events that have started and $T$ the event that have terminated.

**Definition 4.2 (ST-configuration structures)** *An* ST-configuration structure *(also called* ST-structure*) is a tuple $\mathcal{ST} = (ST, l)$ with $ST$ a set of ST-configurations and $l$ a labelling function of the events, $l : \bigcup_{S \in ST} S \to \Sigma$, with $ST$ satisfying the constraint:*

$$\text{if } (S,T) \in ST \text{ then } (S,S) \in ST.$$

The constraint above is a kind of closure so that we do not represent events that are started but never terminated.

**Definition 4.3 (stable ST-structures)** *An ST-configuration structure $(ST, l)$ is called:*

- rooted *iff $(\emptyset, \emptyset) \in ST$;*
- connected *iff $\forall (S,T) \in ST$ non-empty then $\exists e \in S$ s.t. either $(S \setminus \{e\}, T) \in ST$ or $(S, T \setminus \{e\}) \in ST$;*
- closed under bounded unions *iff for $(S,T), (S',T'), (S'',T'') \in ST$ if $(S,T) \cup (S',T') \subseteq (S'',T'')$ then $(S,T) \cup (S',T') \in ST$;*
- closed under bounded intersections *iff for $(S,T), (S',T'), (S'',T'') \in ST$ if $(S,T) \cup (S',T') \subseteq (S'',T'')$ then $(S,T) \cap (S',T') \in ST$.*

*An ST-structure is called* stable *iff it is rooted, connected, and closed under bounded unions and intersections.*

Having only ST-configurations does not give us much information; in particular, we cannot obtain the information that an ST-trace has, cf. [21, def.2.5]. Therefore we define *steps* (or transitions) between ST-configurations. These will give the power to observe the behaviour of a system. We show then how to recover the information than an ST-trace has.

**Definition 4.4 (ST steps)** *A step between two ST-configurations is denoted $(S,T) \xrightarrow{a} (S',T')$ and is defined as:*

**s-step:** $T = T'$, $S \subseteq S'$ and $S' \setminus S = \{e\}$ and $l(e) = a$;

**t-step:** $S = S'$, $T \subseteq T'$ and $T' \setminus T = \{e\}$ and $l(e) = a$.

**Definition 4.5 (paths and traces)** *A* path *of an ST-configuration structure $\mathcal{ST}$ is a sequence of steps and is denoted $\pi$ (overwriting the notation used for paths in HDAs). A path is rooted if it starts in $(\emptyset, \emptyset)$. We generally work with rooted paths; and mention when the discourse involves general paths.*

*For a rooted and connected ST-configuration structure $\mathcal{ST}$ and a path $\pi \in \mathcal{ST}$, the following notion of ST-trace corresponds to the one defined in [21, def.2.5] or [18, sec.7.3]. The ST-trace of $\pi$ is denoted $ST(\pi)$ and is the sequence of labels of the steps of $\pi$ where each label is annotated with $a^0$ if it labels an s-step or by $a^n$ if it labels a t-step, where $n \in \mathbb{N}^+$ is determined by counting the number of steps until the s-step that adds the event $e$ to the $S$ set, with $e$ being the event that has been added to $T$ in the current t-step.*

**Definition 4.6 (adjacent-closure)** *We say that a rooted and connected ST-configuration structure $\mathcal{ST}$ is* adjacent-closed *if the following are respected:*

- *if $(S,T), (S \cup \{e\}, T), (S \cup \{e, e'\}, T) \in \mathcal{ST}$, with $e, e' \notin S \wedge e \neq e'$, then $(S \cup \{e'\}, T) \in \mathcal{ST}$;*

- *if $(S,T), (S \cup \{e\}, T), (S \cup \{e\}, T \cup \{e'\}) \in \mathcal{ST}$, with $e \notin S \wedge e' \notin T \wedge e \neq e'$, then $(S, T \cup \{e'\}) \in \mathcal{ST}$;*

- *if $(S,T), (S, T \cup \{e\}), (S, T \cup \{e, e'\}) \in \mathcal{ST}$, with $e, e' \notin T \wedge e \neq e'$, then $(S, T \cup \{e'\}) \in \mathcal{ST}$.*

**Proposition 4.7 (equivalent with adjacent-closure)** *A rooted and connected ST-configuration structure $\mathcal{ST}$ is called* closed under single events *if it respects the following:*

- *$\forall (S,T) \in \mathcal{ST}$ if $e \in S \setminus T$ then $(S, T \cup \{e\}) \in \mathcal{ST}$;*

- *$\forall (S,T) \in \mathcal{ST}$ if $e \in S \setminus T$ then $(S \setminus \{e\}, T) \in \mathcal{ST}$.*

*A rooted and connected ST-configuration structure $\mathcal{ST}$ that is closed under single events is also adjacent-closed.*

The example of the square with the empty inside is adjacent-closed but not closed under unions nor under intersections. The example of the parallel switch of Winskel [22] is adjacent-closed and closed under unions, but not closed under intersections. The parallel switch can be pictured as only three sides of a cube in *HDA*.

**Proposition 4.8 (correspondence between *HDA* and ST-structures)** *Acyclic and cubical HDAs correspond to rooted, connected, and adjacent-closed ST-structures (denoted as $\mathcal{ST}(\mathcal{H})$).*

**Proof:** Cells correspond to ST-configurations, where the dimension of the cell is given by the number of events present in $S$ but not in $T$. Acyclic should be associated with the fact that we work with sets and each step adds events to one of the two sets $S$ or $T$. The rootedness ensures the existence of the initial cell (with empty sets of events) and the connectedness ensures that each cell of dimension higher than 0 has at least one $s$ map. The special property of a ST-configuration ensures that $s$ and $t$ maps come in pairs. The adjacent-closure then ensures that one cell has all the $s$ and $t$ maps and that cubical laws are respected. The method of "sculpting out" cells from a big cell, or the sticking together of cells by identifying some of their faces, is possible because closure under unions and intersections is not available. Such closures, e.g., would not allow to have an empty square as in the examples before. □

**Correspondence between ST-structures and plain configuration structures:**

Every configuration as in [19, def.5.5] corresponds to an ST-configuration where $S = T$. To every ST-configuration structure $\mathcal{ST}$ we can associate a configuration structure by keeping only those ST-configurations that have $S = T$; i.e., $\mathcal{C}(\mathcal{ST}) = \{T \mid (S,T) \in \mathcal{ST} \wedge S = T\}$.

It is easy to see that if an ST-configuration structure $\mathcal{ST}$ is rooted, or connected, or closed under bounded unions, or intersections, then the corresponding $\mathcal{C}(\mathcal{ST})$ is respectively rooted, connected, closed under bounded unions, or intersections. The rootedness and closure properties are immediate. For the connectedness just apply two times the definition for ST-configurations to obtain that for configuration structures.

But there is not a one to one correspondence between ST-configuration structures and the configurations structures because there can be several ST-structures that have the same configuration structure. The example is of one *HDA* square that is filled in and one that is not; both have the same set of corners and hence the same configuration structure.

For stable ST-structures and stable configuration structures there is a one to one correspondence. One can build from a configuration structure a corresponding ST-structure by adding for each pair of configurations $T$ and $T \cup \{e\}$ the intermediate ST-configuration also; i.e., have the ST-configurations: $(T,T), (T \cup \{e\}, T), (T \cup \{e\}, T \cup \{e\})$ and then close under bounded unions and intersections.

**Definition 4.9 (concurrency and causality)** *For a particular $(S,T)$ of an ST-configuration structure we define the two relations of* concurrency *and* causality *on the events of the ST-configuration as:*

**concurrency:** *is denoted $e||e'$ and defined for $e, e' \in S$ as $e||e'$ iff $\exists (S', T') \subseteq (S,T) : \{e, e'\} \subseteq S' \setminus T'$;*

**causality:** *is denoted $e < e'$ and is defined for $e, e' \in S : e \neq e'$ as $e < e'$ iff $\forall (S', T') \subseteq (S,T) : e' \in S' \Rightarrow e \in T'$.*

Note that on ST-configuration structures the concurrency and causality are not interdefinable, but are disjoint.

The notion of conflicting events (of cancellation as called by Pratt [13]) is not definable for a specific ST-configuration because the notion of cancellation essentially says that the occurrence of one event cancels the others. Conflict/cancellation is a general notion definable on the whole ST-structure.

**Definition 4.10 (conflict)** *For an ST-configuration structure $\mathcal{ST}$ and a set of events $E \in \mathcal{ST}$ the relation of* global conflict *is defined as $\#E$ iff $\nexists (S,T) \in \mathcal{ST}$ with $E \subseteq S$.*

The standard notion of binary conflict is a particular instance of the definition above, where $E = \{e, e'\}$. For stable ST-configuration structures, the constraints of being closed under bounded unions and intersections guarantees that there is no conflict for the events of a particular ST-configuration.

**Definition 4.11 (hh-bisimulation for ST-configuration structures)** *A function $f$ is an* isomorphism *of two ST-configurations $(S,T)f(S',T')$ iff $f$ is an isomorphism of $S$ and $S'$ that agrees on the $T$ and $T'$ sets (i.e., $f \restriction_T = T'$).*

*For two ST-configuration structures $\mathcal{ST}$ and $\mathcal{ST}'$, a relation $R \subseteq \mathcal{ST} \times \mathcal{ST}' \times \mathcal{P}(\mathcal{ST} \times \mathcal{ST}')$ is called a history preserving bisimulation between $\mathcal{ST}$ and $\mathcal{ST}'$ iff $(\emptyset, \emptyset, \emptyset) \in R$ and whenever $((S,T), (S', T'), f) \in R$*

1. *$f$ is an isomorphism between $(S,T)$ and $(S', T')$;*

2. *if $(S,T) \xrightarrow{a} (S_a, T_a)$ then exists $(S'_a, T'_a) \in \mathcal{ST}'$ and $f'$ extending $f$ (i.e., $f' \restriction_{(S,T)} = f$) with $(S', T') \xrightarrow{a} (S'_a, T'_a)$ and $((S_a, T_a), (S'_a, T'_a), f') \in R$;*

3. *if $(S', T') \xrightarrow{a} (S'_a, T'_a)$ in $\mathcal{ST}'$ then exists $(S_a, T_a) \in \mathcal{ST}$ and $f'$ extending $f$ with $(S,T) \xrightarrow{a} (S_a, T_a)$ and $((S_a, T_a), (S'_a, T'_a), f') \in R$.*

*A history preserving bisimulation is called* hereditary *is the following two back conditions hold:*

4. *if* $(S_a, T_a) \xrightarrow{a} (S, T)$ *in* $\mathcal{ST}$ *then exists* $(S'_a, T'_a) \in \mathcal{ST}'$ *and* $f'$ *with* $f\restriction_{(S_a, T_a)} = f'$) *and* $(S'_a, T'_a) \xrightarrow{a} (S', T')$ *and* $((S_a, T_a), (S'_a, T'_a), f') \in R$.

**Proposition 4.12** *For two acyclic and cubical HDAs, $\mathcal{H}$ and $\mathcal{H}'$, their corresponding rooted, connected and adjacent-closed ST-structures $\mathcal{ST}(\mathcal{H})$ and $\mathcal{ST}(\mathcal{H}')$ are hh-bisimilar (cf. Def. 4.11) iff the original higher dimensional automata are hh-bisimilar (cf. Def. 2.7).*

**Proof:** Intuitively, the forward steps in Def. 2.7(1 and 2) are matched in Def. 4.11 by steps (2 and 3). The backward steps in Def. 2.7(5 and 6) are matched by the backward step 4 in Def. 4.11. The $l$-adjacency steps in Def. 2.7(3 and 4) correspond to the restriction 1 in Def. 4.11 of $f$-isomorphism together with the adjacent-closure properties. □

**Proposition 4.13** *For stable ST-structures and their corresponding stable configuration structures the hh-bisimulation from Def. 4.11 corresponds to that of [19, def.9.6].*

The termination predicate of [19, def.9.6] can be defined for ST-structures also, and for *HDAs* using the set of final states/cells. The *hHDML* can also express if these are *maximal* using $\neg\{\}\top$ or $\neg\langle\rangle\top$ to say that no more events can be started or terminated in a cell/ST-configuration.

**Definition 4.14 (*hHDML* interpreted over ST-structures)** *The hHDML logic formulas are interpreted over a ST-configuration structure in a particular ST-configuration. The during modalities $\{\}$ and $\overline{\{\}}$ are moving on the s-steps, forward respectively backward; whereas the after modalities $\langle\rangle$ and $\overline{\langle\rangle}$ move on the t-steps. The rest is the same as in Def. 3.2.*

Intuitively, when *hHDML* is interpreted in an ST-configuration $(S, T)$ a formula $\{a\}\varphi$ says that one new event labelled with $a$ should be added to the set of started events $S$ and the resulting ST-configuration should be part of the ST-structure we are working with and the formula $\varphi$ should hold.

**Proposition 4.15** ($\overset{hHDML}{\sim}$ **and** $\overset{hh}{\sim}$ **coincide over ST-structures**) *For rooted, connected, and adjacent-closed ST-configuration structures, the relations $\overset{hHDML}{\sim}$ (Def. 3.3) and $\overset{hh}{\sim}$ (Def. 4.11) coincide.*

**Proof:** The more difficult part is to show that $\overset{hHDML}{\sim}$ satisfies the restrictions of being a $\overset{hh}{\sim}$ over ST-configuration structures from Def. 4.11. The changing in the map indexes, or equivalently in interchanging cell associations, that we were doing in Prop. 3.4 is here reflected in the change of the isomorphism $f$ that the $\overset{hh}{\sim}$ is defined with. The isomorphism is changed by interchanging event associations. We need to change the isomorphism when problems appear from the adjacent-closure property of the ST-structures that we work with. This change is always possible, without braking the other forward and backward properties of the $\overset{hh}{\sim}$, and such that it also caters for the adjacent-closure property. □

The notion of *concurrent step* [19, def.7.1] can be defined over ST-configuration structures (or *HDAs*) and captured in the *hHDML* logic by restricting the language of the logic to using only syntactic definitions of the form $\langle A\rangle\varphi$ interpreted in the states (cells of dimension 0) of the *HDAs*.

The syntactic definition for a multiset of labels $A$ is $\langle A \rangle \triangleq \{A\}\langle A \rangle \varphi$ where $\{A\}$ is $\{a_1\}\ldots\{a_n\}$ with $a_i \in A$ (analogous for $\langle A \rangle$). The concurrent steps of [19, def.7.1] become just $(S,T) \xrightarrow{A} (S',T')$ with $T' = T$ and $S' = S \cup \{a_1, \ldots, a_n\}$ for $a_i \in A$, if the ST-configuration $(S', T')$ is reachable from $(S,T)$ through a sequence of only s-steps. The standard Hennessy-Milner logic formulas and the transitions in labelled transition systems are the restriction of concurrent steps and formulas from above to $A$ being a singleton set.

## 5  Conclusion

We presented the history-aware higher dimensional modal logic as a response to the question of what is a minimal extension in the style of standard modal logic of the previous *HDML*, with a natural interpretation over the higher dimensional automata, that can capture the hereditary history preserving bisimulation defined for this model of concurrency. This logic with past modalities does not employ event variables (opposed to [1,7]), but it uses the new modalities that talk about what happens *during* the concurrent execution of events (besides the standard *after* modalities).

In the second part of the paper we have introduced the *ST-configuration structures* as a model of concurrency extending the configuration structures of [20] to the setting of *HDAs*. We have given various related definitions and made correlations with the similar notions from the other models that we relate these with, i.e., (stable) configuration structures and (acyclic and cubical) *HDAs*. The new *hHDML* was interpreted over these and the result of capturing hh-bisimulation was shown again in this new setting.

**Acknowledgements:**

# References

[1] P. Baldan and S. Crafa. A Logic for True Concurrency. In *CONCUR'10*, volume 6269 of *LNCS*, pages 147–161. Springer, 2010.

[2] P. Blackburn, M. de Rijke, and Y. Venema. *Modal Logic*, volume 53 of *Cambridge Tracts in Theor. Comput. Sci.* Cambridge Univ. Press, 2001.

[3] V. Gupta. *Chu Spaces: A Model of Concurrency*. PhD thesis, Stanford University, 1994.

[4] D. Harel, D. Kozen, and J. Tiuryn. *Dynamic Logic*. MIT Press, 2000.

[5] F. Laroussinie, S. Pinchinat, and P. Schnoebelen. Translations Between Modal Logics of Reactive Systems. *Theor. Comput. Sci.*, 140(1):53–71, 1995.

[6] O. Lichtenstein, A. Pnueli, and L. D. Zuck. The Glory of the Past. In *Conference on Logics of Programs*, volume 193 of *LNCS*, pages 196–218. Springer, 1985.

[7] I. Phillips and I. Ulidowski. A Logic with Reverse Modalities for History-preserving Bisimulations. In *EXPRESS'11*, volume 64 of *EPTCS*, pages 104–118, 2011.

[8] I. Phillips and I. Ulidowski. A Hierarchy of Reverse Bisimulations on Stable Configuration Structures. *Math. Struct. in Comp. Science*, 22:333–372, 2012.

[9] V. R. Pratt. Modeling concurrency with geometry. In *POPL'91*, pages 311–322, 1991.

[10] V. R. Pratt. Chu spaces and their interpretation as concurrent objects. In *Computer Science Today: Recent Trends and Developments*, volume 1000 of *LNCS*, pages 392–405. Springer, 1995.

[11] V. R. Pratt. Higher dimensional automata revisited. *Math. Struct. Comput. Sci.*, 10(4):525–548, 2000.

[12] V. R. Pratt. Event-State Duality: The Enriched Case. In *CONCUR'02*, volume 2421 of *LNCS*, pages 41–56. Springer, 2002.

[13] V. R. Pratt. Transition and Cancellation in Concurrency and Branching Time. *Math. Struct. Comput. Sci.*, 13(4):485–529, 2003.

[14] C. Prisacariu. Modal Logic over Higher Dimensional Automata. In *CONCUR'10*, volume 6269 of *LNCS*, pages 494–508. Springer, 2010.

[15] C. Prisacariu. Higher Dimensional Modal Logic. 2011. (journal draft under review; available at http://heim.ifi.uio.no/~cristi/publications.shtml).

[16] C. Prisacariu and G. Schneider. The Glory of the Past and Geometrical Concurrency – technicalities. Technical report, Department of Informatics, University of Oslo, 2012.

[17] R. J. Van Glabbeek. The Linear Time – Branching Time Spectrum I. In *Handbook of Process Algebra*, pages 3–99. Elsevier, 2001.

[18] R. J. van Glabbeek. On the Expressiveness of Higher Dimensional Automata. *Theor. Comput. Sci.*, 356(3):265–290, 2006.

[19] R. J. van Glabbeek and U. Goltz. Refinement of actions and equivalence notions for concurrent systems. *Acta Informatica*, 37(4/5):229–327, 2001.

[20] R. J. van Glabbeek and G. D. Plotkin. Configuration structures, event structures and Petri nets. *Theor. Comput. Sci.*, 410(41):4111–4159, 2009.

[21] R. J. van Glabbeek and F. W. Vaandrager. The Difference between Splitting in n and n+1. *Information and Computation*, 136(2):109–142, 1997.

[22] G. Winskel. Event structures. In *Advances in Petri Nets*, volume 255 of *LNCS*, pages 325–392. Springer, 1986.